# Does built environment influence physical activity and body postures in homework journeys?

**Isaac Debache**
  Institut pluridisciplinaire Hubert Curien

**Cédric Sueur** ( cedric.sueur@iphc.cnrs.fr )
  Institut pluridisciplinaire Hubert Curien   https://orcid.org/0000-0001-8206-2739

**Julie Vallée**
  Geographie-cites

**Audrey Bergouignan**
  Institut pluridisciplinaire Hubert Curien

**Basile Chaix**
  Institut Pierre Louis d'Epidemiologie et de Sante Publique

# Abstract

Understanding the effects of built environment on physical activity is important for promoting healthy lifestyles in cities. Yet, very few studies have used objective continuous location data and measures of physical activity while addressing biases to causal inference. In addition, the effect of the environment on body postures during trips, albeit of physiological importance, is rarely addressed. Using mixed models for compositional data on sensor-derived physical activity information, we estimated the effects of greenery, destination density, public transport time efficiency, and average area education among 692 home-work journeys made by 121 healthy adults (80 men, 41 women). Higher levels of greenery, public transport time efficiency, and average area education along the shortest network path between home and workplaces were found to reduce contemporaneous sedentary postures and increase physical activity. These relationships suggest that decision makers should consider greening cities and improving public transport efficiency as an effective way to reduce the prevalence of sedentary behaviors in our societies.

# Introduction

Because of its well-known effects on reducing mortality and health hazards, such as cardiovascular and coronary diseases, diabetes, cancer, and depression, physical activity (PA) is widely promoted by public health policies (1,2). Despite these efforts, the prevalence of physical inactivity (lack of PA) has not declined over the last decade, while the time spent in sedentary behaviors (SB, i.e., sitting or reclining with an energy expenditure below 1.5 Metabolic Equivalents (3)) has kept increasing (4). As a potential field of intervention, research has studied various features of social and physical environments and their link to PA. To date, despite considerable discrepancies in the literature, there is some evidence for the positive effect of certain environmental features, such as the presence of green spaces and paths, population and destination density, access to transit and walkability, on both leisure PA and active transportation (5–7). The goal of the present study is to bring novel, accurate evidence on the relationship between four key features of the physical and social environments (greenery level, destination density, access to the public transportation network and average neighborhood education) and PA behaviors. This is achieved by proposing a new study design a) using objective, contemporaneous measures of environmental features and PA, b) aiming to improve causal inference through a sound definition of environmental exposures, and c) distinguishing between objective measures of PA and SB, i.e. between body motion and body posture.

### A contemporaneous design

The largest part of the evidence regarding the effect of environments on PA was obtained by linking environmental attributes of a fixed location, such as the subjects' home or workplace, with PA levels aggregated over a certain period (e.g., a day or a week). However, researchers have argued that, individuals being mobile, their PA at a certain moment should also be regarded as a function of their contemporaneous exposure to environment (8–10). Following recent studies, we implemented this design (hereafter 'contemporaneous design') by assessing environmental features taking into account the current location of subjects and their simultaneous accelerometer-derived PA (11,12). With this method, the close time correspondence between the exposure and the outcome yields accurate and more valid estimates of the associations between the two.

### Selection biases and causal inference

Studies on the association between environments and PA are subject to two well-identified methodological biases among others. First, the *residential self-selection* bias refers to the preference of individuals inclined to engage in PA for neighborhoods of residence favorable to PA (5). To address the residential self-selection bias, we considered subjects' reported motivation in choosing their residential neighborhood. Second, the *selective daily mobility* bias, common in studies implementing the contemporaneous design, refers in our context to the fact that the preferences of participants influence both their momentary locations and their physical activity at these locations. Despite the growing number of studies with a contemporaneous design, the selective mobility bias has rarely been addressed in previous studies (10,13,14). We adopt two approaches to address this bias in our study.

First, rather than calculating environmental exposures along the actual observed GPS path, we measured environmental exposures along the shortest recalculated home-work paths. Considering actual itineraries lead to bias because people tend to modulate the itinerary that they chose according to their transport mode (10). In contrast, we assume that the characteristics of the shortest route causally influence the transport mode that individuals will use for their trip.

Second, *trip purpose* is usually omitted from the models, although it affects both the exposure to environmental features and the travel mode (15–18). For instance, trips to shopping malls might require the use of motorized transport to carry the purchases, while being characterized by a high exposure to a specific land-use environment. To address this second source of selective daily bias, we limited our frame of analysis to the sole utilitarian journeys between home and workplace. Indeed, reducing the sample to only one trip purpose removes confounding from this variable.

Thus, the present study builds on the advantage of a contemporaneous design while minimizing biases that compromise causal inference. In addition, irrespective of issues of causality, home-work journeys are worth investigating as an opportunity to achieve the daily recommended PA (19) , since they make up a considerable share of the daily travel time in the working population and can involve substantial episodes of PA.

### Physical activity and body postures

Past studies using objective measurements focused only on PA (typically walking or cycling), but not on SB (reclining/sitting postures). Yet, with recent findings regarding the deleterious effects of prolonged SB (20–23), the focus of research has shifted towards a refined definition of PA, including not only body motion, but also the postures (24,25). In the present study, we investigated the effect of the environment on both PA and SB, by considering different categories of physical behaviors, such as sitting/reclining, standing, or walking.

### Environmental influences

In a literature review, Kondo (26) identified 12 articles using objective location and PA data that found a positive association between exposure to greenery and contemporaneous levels of moderate-to-vigorous PA. Yet, as we pointed out, the validity of such studies is mitigated by selective mobility bias (10). Notwithstanding this limitation, we rely on these findings to posit that the observed human preference for green environments when engaging in PA should result in more active commuting along routes featuring greenery.

Likewise, accessibility to destinations required for daily living, such as shops, grocery markets etc. is as a factor that promotes walking for transports (27). Yet, despite a large body of evidence (28–31), causal evidence from a contemporaneous design is still missing. Here, we hypothesized that  high destination density represents a rich, stimulating environment, which encourages active commuting (10).

Overall, better access to public transport system is thought to have a positive impact on PA levels, as travelling with public transports typically implies more walking episodes than travelling with a private car (32–35). Thus, we hypothesized that greater incentives of commuting with public transports, such as better access and higher time efficiency, would also be associated with more PA and less SB over the journey. SB is of particular interest here, as public transports compared to private cars often require commuters to be standing.

According to the theory of environmental inequality, low socio-economic neighborhoods are less likely to benefit from health-promoting environmental factors, such as clean spaces with well-maintained buildings, and more likely to be exposed to health-damaging ones, such as sources of pollution or crime (36,37).  Some studies based on subjects' self-reported PA showed an increase in utilitarian walking in high education neighborhood, independent of individual socio-economic status (SES) (38,39). Here, we posit that area SES is a proxy to a variety of unknown environmental factors related to the pleasantness of the environment affecting PA, and therefore expect that work-home routes crossing high SES neighborhoods, as captured by the residents' average area educational attainment, should feature more PA and less SB. In addition, we regard area education as a proxy for environmental confounding that could bias the estimates for the effects of other variables of interest, and add it systematically to the linear models.

## Material And Methods

### Population

The present study uses data collected in the RECORD (Residential Environment and COronary heart Disease) study, which investigated spatial disparities in health. From February 2007 to March 2008, individuals who came to one of the four sites of the IPC (Investigation Préventive et Clinique) Medical Center for a free medical examination offered by the French National Insurance System for working, unemployed, or retired employees and their families were invited to enter the RECORD study. Eligibility criteria were age 30-79 years and residence in ten given districts of Paris (out of 20) or in 111 other municipalities of the Ile-de-France region, as well as sufficient cognitive and linguistic abilities to comply with the guidelines. During the second wave of the study (between September 2013 and June 2015), former and newly recruited participants underwent a medical examination, after which they were invited to enter the RECORD MultiSensor ancillary study whenever sensors were available. In this study, they were asked to wear body-mounted sensors, including a pair of accelerometers and a GPS receiver. Participants in this ancillary study were instructed to wear the sensors for 7 consecutive days, as they carried out their usual activities in free-living conditions, and to keep a logbook with the places that they had visited, as well as the transport modes used in journeys between those places. Participants' trips during the monitoring period were determined by combining processed GPS tracks with a mobility survey conducted through the phone (the method is described in detail elsewhere (35)). In addition, participants were asked to answer a questionnaire regarding their health and dietary habits, neighborhood, demographics, and SES. The study protocol was approved by the French Data Protection Authority (Decision No. DR-2013-568 on 2/12/2013), and a written informed consent was filled by all participants.

The present study only retained home-work trips in the data. We removed journeys that included segments outside of Ile-de-France (region of Paris). Journeys that included any stop not related to travel (for example stopping at shops or at friends' house) were not included in the analysis. The final dataset included 692 journeys recorded for 121 participants.

### Physical behaviors during travel

Direct measures of PA and SB were derived from the two tri-axial Vitamove Research-V1000® devices, worn at the right upper leg and on the chest during wake time (except for water-based activities). From the acceleration signal, the VitaScore software derives motion type and orientation of the body compared to the gravitational axis, which are combined to determine the subjects' behavior. We grouped the behavioral categories provided by the software into three broader categories: SB (lying or sitting), standing (ST) still and light movements and PA (walking, running and bicycling).

### Environmental exposures

The environmental attributes of the journeys were calculated along the shortest (walking) routes between subjects' homes and workplaces, which was determined using GoogleMaps.

*Greenery index*

We used two methods to measure greenery level. First, we sampled a set of equidistant points (each 200 meters) along the shortest routes and calculated the mean of the shortest distances between the points and the network of green paths. This variable represents the opportunity cost, in terms of distance covered, of using the green network. Second, we calculated the proportion of green spaces for a buffer zone of 100m radius around the shortest routes, capturing the direct proximity to greenery during the journey, as illustrated in Figure 1. We calculated these measures using the 2008 open data of the Ile-de-France Institute for Urbanism (40).

*Destination density*

Destination density was calculated as the number of destinations per km$^2$, including public services, shops, entertainment facilities etc., in a buffer zone of 100m radius around the shortest routes (French National Institute for Statistics and Economic Studies INSEE, 2011 (41)).

*Accessibility to and time cost of public transportation*

Two methods were used to estimate the incentive to use public transports. First, for the residence and workplace, the Euclidian distance to the nearest bus, metro, tramway, or railway station (Île-de-France Mobilités, 2012 (42)) was calculated, and the larger distance of the two was defined as the accessibility variable. Second, the time cost variable was defined as the ratio of the travel time using public transports to the travel time using a car, as estimated by GoogleMaps.

*Area education*

To estimate the SES of the areas crossed by the participants, we used the buffers of 100m radius around the shortest route for the journey. In each buffer, the educational attainment of the population was defined as the share of residents aged 20 and over holding a university degree (INSEE 2010 census, geocoded at the street address level (43)).

### Other covariables

Several potentially confounding factors known to influence PA were accounted for. At the journey level, the length (in km) of the shortest route was accounted for as it probably plays a role in the choice of the travel mode. At the individual level, we considered sex, age, being in couple, having children under 14 years at home, household income (by tertiles), and individual educational attainment (no higher education, undergraduate, graduate). To control for neighborhood selection bias, we used questionnaire data (17) where participants were asked to score, on a scale varying from 'not at all' to 'very much' (coded 0-3), how important the greenery level, presence of shopping facilities, SES, and accessibility of public transports were in the choice of neighborhood to which they moved.

### Statistical analysis

The goal of the analysis was to estimate the effect of exposure to environmental route attributes on the time budget of physical behaviour during the journey. Thus, the physical behaviours were regarded as compositional data adding up to 1, where each of the three parts corresponds to the share of the time spent in sedentary postures, standing, and PA (25). Compositions add up to a constant sum and are hence interdependent and constrained between 0 and 1. To model compositions of physical behaviors ( ) as dependent variables explained by environmental measures, we applied the additive log ratio (alr) transformation to the data (44), taking sedentary time as reference:

$$\mathbf{y}' = [0, \ln \frac{y_{ST}}{y_{SB}}, \ln \frac{y_{PA}}{y_{SB}}].$$

With this linear transformation, the log-ratios can take any real value and be modelled using usual tools of statistical analysis, while preserving their relational nature.

To avoid infinite values when applying the alr transformation an epsilon (0.01) was added to all parts. The two log-ratios were modelled as linear functions of the environmental and social factors and co-variables, using mixed linear regression with participants as random intercepts. The predicted log-ratio vector was then back-transformed to predict a composition for any set of values of environmental features.

The environmental factors (greenery ratio, distance to greenery, destination density, time cost of public transport, distance to the nearest station, and average area education,) were included with the following co-variables: age, sex, being in couple, having children at home, individual educational level, income level, length of shortest route, squared length, and neighborhood selection factors. A squared term was added for route length as we assumed that it has a non-linear association with the probability to be active. In all models, average area education was included as a proxy for the SES of the areas crossed, as we suspected it to be a confounding variable causally influencing both the environmental explanatory variables and the outcome of interest. As we had no reason to assume that the relationships between environmental factors and physical behaviors were linear, we tested a squared term and retained it in the models only if relevant (p-value<0.05 for a likelihood ratio test).

All analyses were run using R (45), with libraries 'rgeos' (46) 'sf' and 'sp' (47) for spatial analysis and 'lme4' for mixed models (48).

## Results

### Description of participants

The 121 participants in the study were in average 59 years old (standard deviation = 8 years), and 66% of them were men. As explained elsewhere (25), the population had a slightly higher SES than the French average (income and education). Descriptive statistics of the population are shown in Table 1.

### Description of journeys

The median length of the 692 journeys (based on GPS data) was of 7.64 km [1st and 9th deciles: 1.20, 23.3], and their median duration was of 27.9 minutes [12.0, 52.9]. On average, participants spent 49.9% [0%, 94.6%] of the travel time in SB (typically sitting or reclining), 16.6% [2.6%, 62.9%] standing or performing light body movements (ST), and 19.6% [0%, 72.4%] performing PA (typically walking and bicycling). Full descriptive statistics of all variables used in the models are shown in Table 1.

For descriptive purposes, we examined the distribution of physical behaviors by different travel modes as assessed by the GPS-based mobility survey: journeys that are entirely active (N=112), journeys including public transports (N=307), and other journeys (i.e., including some private cars, N=273). Entirely active journeys were, on average (i.e. closed geometric mean), composed of large shares of PA (65%), some ST (28%) and very little SB (7%). In journeys including public transports, compositions of physical behaviors were more balanced (PA: 32%, ST: 28%, SB: 40%). Other journeys were mostly sedentary (PA: 6%, ST: 22%, SB: 73%). The compositions of physical behaviors over the journeys by commute mode are shown in a ternary plot (Figure 2).

### Model results

Results of the models are presented as ST/SB ratios or PA/SB ratios (followed by the 95% confidence interval in square brackets) throughout this section, and in Table 2. As ratios are less intuitive, we added the predicted compositions for the range of values taken by the explanatory variables of interest. These are reported in Table 3 and illustrated in Figure 3. In the next sections, we first report associations from models that included only one environmental factor (either greenery, destinations, or public transport), adjusted for area SES and individual confounders.

*Greenery*

The proportion of area along the shortest route covered with green spaces predicted a more active composition of physical behaviors over the journey. A 0.01 increase in the proportion of green spaces led to a 11% [+2%, +22%] increase in the ST/SB ratio, and a 13% [3%, 23%] increase in PA/SB. The effect on time-budgets of PA behaviors is illustrated in Figure 3. It should be noted that the green ratio for the journeys was typically bound between 0.01 and 0.07 (1$^{st}$ and 9$^{th}$ deciles).

*Destination density*

Models revealed that a 1 SD (standard deviation) increase in destination density (+695 destinations per km$^2$) was not associated with either the ST/SB ratio (-4% [-33%, +37%]) or the PA/SB ratio (+18% [-17%, 70%]).

*Public transports efficiency*

We observed essentially no association between a 1 SD increase in the maximum distance to public transport stations (70m) and the ST/SB [-5% (-27%, 23%)] and PA/SB ratios [+16% (-10%, 50%)]. However, the public transit to car time ratio had a positive association with SB. One SD increase in this comparative time cost (+80% in travel duration using public transports compared to car) had an estimated association of -20% [-36%, +0%] with the ST/SB ratio and –26% [-40%, -8%] with the PA/SB ratio.

*Average area education*

Among the variables of interests, average area education along the shortest route was the only variable to be non-linearly associated with behavior. At the lowest level of area education, a 1% increase in area education was associated with an increase of the ST/SB ratio of +18% [1%, +37%], and the PA/ST ratio of +26% [+8%, +46%]. Due to negative quadratic coefficients, the relationship between area education and ST or PA follows an inverted J-shape pattern, reaching its maximum around the mean education score (52%). As the distribution of the percentage of high-educated people is right-skewed (1$^{st}$ decile = 30%, 9$^{th}$ decile = 62%), the positive marginal effect on ST and PA for low levels of area education is larger than the negative marginal effect at high levels of education.

*Other variables*

The socio-demographic variables included in the models did not exhibit any clear associations with the PA behaviors. In contrast, as expected, the length of the shortest route was clearly associated with an increase in SB.

*Assessing the independent effect of the variables of interest*

Table 4 reports the results of a model including all variables of interest that were found to be relevant in separate models for the time budget of physical activity (greenery index, mean area education, and transport time cost). These do not point to any substantial difference between the models, and suggest that these associations are independent.

# Discussion

*Interpretation of findings*

In this study, we investigated the effects of environmental attributes of home-work routes on physical behaviors during the journeys in urban adults. It yielded three important insights. First, high proportions of green area along the route increased the share of time spent in non-sedentary behaviors (standing or being physically active). Second, higher time cost of public transport compared to private car resulted in lower shares of time spent standing and in PA. Third, a higher residents' level of education along the route was associated with more standing and PA in the lower range of the education variable, but the relationship was reversed (although weaker) in the higher range of this variable.

Our finding about the positive association between greenery and PA agrees with some of the previous work on this topic (26). However, to our knowledge, our study is the first to assess this relationship in a contemporaneous design with sensor-derived activity and posture data.

This association was better captured when considering all green spaces intersecting the buffers around the routes than when considering only the spatial accessibility to green paths. This can point out to a visual sensitivity to greenery in the environment present within sight range (100m), rather than to the very accessibility to green path.

A lower time cost of public transport was associated with active behaviors, in accordance with previous literature (32–34). Distance to public transport stations, however, did not clearly predict any behavior. It should be noted that, for our population, 90% of the trips started and ended within 255m of a station, meaning that the distance to the station is unlikely to play a major role in the choice of commute mode. Indeed, it should be emphasized that the present study was conducted in the population-dense region of Paris, which benefits from a well-developed network of public transport, and that results should be interpreted while keeping this specific context in mind. In addition, a longer distance to the closest station might have conflicting effects, as it is an incentive to use car and thereby to reduce activity, but it increases activity when actually walking to the station.

Area education, up to a certain level, had a negative effect on SB and a positive one on PA, independent of personal SES and other environmental variables tested here. This finding concurs with a previous report for an adult, urban French Canadian population (38) showing increased levels of PA in well-off neighborhoods compared to socially deprived neighborhoods. It is plausible that area education is a proxy for a large set of environmental features related to the agreeableness of the environment, both social and physical, which favor physical activity. Regarding why this relationship was attenuated and slightly inverted in the highest quantiles of area education, future research could examine whether it is attributable to other relevant variables that were not examined here.

Our hypothesis regarding the positive effect of destination density on PA, observed in several previous studies (28–31), could not be verified. Destination-rich routes can be an incentive to walk or bicycle because they offer good opportunities to visit places, such as shops, along the way, and also because they offer a visually rich and pleasant environment. In fact, considering all sorts of travels in the same population, Chaix et al. found a positive association between destination density and activity (49). The fact that we focused here on direct home-work trips (omitting journeys including stops that are not travel-related) is therefore a potential explanation for the absence of association with destination density. Finally, it is interesting to note that our models suggest that people who report giving importance to the presence of shops in their neighborhood are much likelier to be non-sedentary in journeys.

*Strengths and limitations*

By looking at the home-to-work and work-to-home journeys, the present study was able to reduce heterogeneity in the study of environmental effects on physical behaviors during trips and neutralize the confounding effect of trip purpose. Moreover, defining environmental exposures along recalculated, shortest-path itineraries rather than along actual paths allowed us to address a major issue that would undermine causal inference. More than assessing the co-occurrence of environmental exposures and PA, it aimed at determining the causal effects of environments on physical behavior, using precise, sensor-derived data. In addition, this study is one of the very few to address the question of posture during commuting using objective measurements of body posture.

Yet, the issue of neighborhood selection is not fully addressed in this study: we still could not fully control for whether a specific workplace was chosen by an individual because it fitted her/his preferred commute mode. Thus, it would have been useful to ask participants whether the concordance between the environment around their workplace and their preferred commute mode played a role in the choice of a workplace, as we did for the choice of the residence. However, we argue that such considerations are very unlikely to be critical in the choice of a workplace, especially after we controlled for factors related the choice of residence.

In this study, the shortest path was determined using GoogleMaps (2019). With Google Maps, it was impossible to calculate itineraries back to the time of the study. We therefore had to assume that changes that occurred in the topography of traffic infrastructures in the Paris region over the last 4-6 years did not greatly alter the routes taken by individuals and that, if it were the case, it would not substantially affect the levels of exposure to our variables of interest. The same caveat applies to our environmental data, which were few years older than the study period.

The analytical framework proposed here models the effect of environmental attributes on time *proportions* of physical activity behaviors, and not on absolute time. We made this choice since we focus on home-work journeys as an autonomous daily life-segment of varying duration. However, one can argue that a supposed environmental effect might differ depending on the absolute journey duration. For instance, a green, activity-generating environment could have a stronger impact on the decision to walk 20 minutes in a 40-minute journey than on the decision to walk 2 minutes in a 4-minute journey. Nevertheless, since this caveat seemed less relevant within the observed range of journey durations (1$^{st}$ decile = 12 minutes, 9$^{th}$ decile = 53 minutes), we have favored a compositional approach that is independent of the absolute duration (while controlling for journey length).

Last, the present study does not account for the time of the day at which the journey was made. Depending on the time of day, effects of the environmental features studied here may vary: for instance, greenery might be less enjoyable at night, and the efficiency of public transports less relevant during low-traffic hours (50). Hourly variability in area attributes, i.e., the 'daycourse of place' (50), should be investigated in future research to better understand the role of environment in physical activity practice.

## Conclusion

Using objective, precise sensor-derived measures and an innovative design attempting to maximize the quality of causal inference, this study generates new insights bridging urban planning and public health. Individual variables and journey length being controlled for, individuals were less likely to be sedentary during journeys across areas featuring more greenery. Likewise, they were less likely to be sedentary during journeys that could be made with time-efficient public transports. Results also pointed to important disparities in the levels of PA across neighborhood SES, with the highest level of PA achieved when the routes for commute crossed middle-to-high SES neighborhoods. We conclude that developing greenery and time-efficient public transport networks are urban planning steps that can effectively help fighting the pandemic of physical inactivity and sedentary behavior, and that attention should be paid to equal distribution of activity-generating environmental features across neighborhoods.

## Tables

Table 1: **Descriptive statistics of variables in the study (N = 121)**

|  |  | Quantiles |  |  |  |  | Mean | Standard deviation |
|---|---|---|---|---|---|---|---|---|
|  |  | 0% | 10% | 50% | 90% | 100% |  |  |
| Personal variables | age (years) | 0 | 49 | 60 | 69 | 82 | 59 | 8 |
|  | sex (male=1) |  |  |  |  |  | 0.66 |  |
|  | couple |  |  |  |  |  | 0.71 |  |
|  | children<14 y. |  |  |  |  |  | 0.45 |  |
|  | income category | 1 | 1 | 2 | 3 | 3 | 1.83 | 0.86 |
|  | educational category[1] | 1 | 1 | 3 | 3 | 3 | 2.27 | 0.91 |
| Reported importance of variable in choice of neighborhood of residence | greenery | 0 | 0 | 2 | 3 | 3 | 2.06 | 1.00 |
|  | shops | 0 | 1 | 2 | 3 | 3 | 2.24 | 0.86 |
|  | SES | 0 | 0 | 2 | 3 | 3 | 1.50 | 0.97 |
|  | public transports | 0 | 0 | 3 | 3 | 3 | 2.26 | 1.07 |
| Trip attributes | Length of actual itinerary (km) | 0.04 | 1.20 | 7.64 | 23.33 | 64.33 | 10.66 | 9.95 |
|  | duration (min) | 4.00 | 12.04 | 27.85 | 52.91 | 243.00 | 31.61 | 19.34 |
|  | green ratio | 0 | 0.0095 | 0.0336 | 0.0715 | 0.1389 | 0.0382 | 0.0274 |
|  | green distance (m) | 42 | 137 | 341 | 606 | 1919 | 360 | 198 |
|  | destination density ($km^{-2}$) | 4 | 155 | 487 | 1854 | 4476 | 777 | 695 |
|  | mean area education | 0.1785 | 0.2958 | 0.5149 | 0.6189 | 0.7026 | 0.4854 | 0.1261 |
|  | transp. proximity (m) | 31 | 91 | 152 | 255 | 532 | 162 | 70 |
|  | trans. time cost | 0.81 | 1.26 | 1.84 | 2.89 | 7.51 | 2.01 | 0.80 |
| Composiiton of physical behaviors | sedentary | 0 | 0 | 0.4988 | 0.9456 | 1 | 0.474 | 0.3572 |
|  | standing | 0 | 0.0258 | 0.1654 | 0.6291 | 1 | 0.2554 | 0.252 |
|  | PA | 0 | 0 | 0.1957 | 0.7244 | 1 | 0.2705 | 0.2718 |
| Commute mode | active |  |  |  |  |  | 0.16 |  |
|  | public transports |  |  |  |  |  | 0.53 |  |

[1] 1 = no higher education, 2 = undergraduate, 3 = graduate

Table 2: **Exponentiated coefficients (β) [95% confidence interval] of mixed linear regressions** modelling the relationship between personal and environmental variables and the log ratios ST/SB and PA/SB. After a change of one unit in the independent variable, the ratio is estimated as β times the ratio before the change.

| Variable of interest | Green ratio (%) | | Distance from green (z-score) | | Destination density (z-score) | |
|---|---|---|---|---|---|---|
| | ST/SB | ST/PA | ST/SB | ST/PA | ST/SB | ST/PA |
| Intercept | 0.0043 | 0.0008 | 0.0075 | 0.0015 | 0.0016 | 0.0004 |
| | [0.0001, 0.1574] | [0, 0.0298] | [0.0002, 0.2553] | [0, 0.052] | [0, 0.0545] | [0, 0.0127] |
| Variable of interest | 1.1125 | 1.1292 | 0.8104 | 0.8092 | 0.9597 | 1.1853 |
| | [1.016, 1.2181] | [1.0324, 1.235] | [0.6253, 1.0503] | [0.6265, 1.0452] | [0.6739, 1.3666] | [0.8265, 1.6998] |
| Reported importance of var. of interest | 0.9986 | 0.9006 | 0.9968 | 0.9042 | 1.7746 | 1.7151 |
| | [0.6966, 1.4316] | [0.6181, 1.3122] | [0.6974, 1.4247] | [0.6203, 1.3182] | [1.1905, 2.6453] | [1.124, 2.617] |
| Age (years) | 0.9484 | 1.0357 | 0.954 | 1.0452 | 1.0046 | 1.1001 |
| | [0.655, 1.3733] | [0.7033, 1.5252] | [0.6605, 1.3777] | [0.709, 1.5408] | [0.7025, 1.4367] | [0.7526, 1.6081] |
| Sex (1 = male) | 0.6786 | 0.7731 | 0.7113 | 0.8203 | 0.8703 | 1.0549 |
| | [0.3143, 1.4651] | [0.3461, 1.7266] | [0.3317, 1.5252] | [0.367, 1.8336] | [0.4155, 1.8229] | [0.4822, 2.308] |
| Income category = 2 | 0.5835 | 0.5413 | 0.5806 | 0.5411 | 0.6637 | 0.6705 |
| | [0.2361, 1.4421] | [0.2105, 1.3921] | [0.2366, 1.4245] | [0.21, 1.3939] | [0.2709, 1.6256] | [0.2601, 1.7288] |
| Income category = 3 | 0.6806 | 0.6083 | 0.7201 | 0.6509 | 0.8683 | 0.8712 |
| | [0.26, 1.7814] | [0.2227, 1.6615] | [0.2779, 1.8656] | [0.2384, 1.7769] | [0.3329, 2.2645] | [0.316, 2.4023] |
| Education category = 2 | 2.1851 | 2.9633 | 2.5532 | 3.4898 | 1.934 | 2.4559 |
| | [0.6586, 7.2498] | [0.8465, 10.3738] | [0.7724, 8.4397] | [0.9892, 12.3114] | [0.5952, 6.2841] | [0.705, 8.556] |
| Education category = 3 | 1.2159 | 1.698 | 1.2674 | 1.7772 | 1.2471 | 1.7323 |
| | [0.5198, 2.8442] | [0.6987, 4.1267] | [0.5461, 2.941] | [0.7309, 4.3213] | [0.5461, 2.848] | [0.722, 4.1564] |
| Couple (1 = yes) | 0.7915 | 0.5818 | 0.8215 | 0.6074 | 0.7326 | 0.5682 |
| | [0.3346, 1.8724] | [0.2366, 1.431] | [0.3501, 1.9278] | [0.2468, 1.4947] | [0.3171, 1.6928] | [0.234, 1.3798] |
| Children <14 at home (1 = yes) | 1.5211 | 1.9074 | 1.4945 | 1.8633 | 1.2129 | 1.4405 |
| | [0.6845, 3.3804] | [0.8277, 4.3955] | [0.6774, 3.297] | [0.808, 4.2969] | [0.5604, 2.6251] | [0.635, 3.2679] |
| Mean area education (%) | 1.2138 | 1.2926 | 1.2005 | 1.2765 | 1.2158 | 1.2895 |
| | [1.0421, 1.4138] | [1.1115, 1.5032] | [1.0318, 1.3968] | [1.0977, 1.4844] | [1.0471, 1.4116] | [1.1107, 1.497] |
| Squared mean area education (%) | 0.9981 | 0.9975 | 0.9982 | 0.9976 | 0.9981 | 0.9974 |
| | [0.9964, 0.9998] | [0.9958, 0.9991] | [0.9966, 0.9999] | [0.9959, 0.9993] | [0.9964, 0.9997] | [0.9957, 0.999] |
| Length (km) | 0.2441 | 0.1247 | 0.2473 | 0.1249 | 0.2444 | 0.1293 |
| | [0.1659, 0.3592] | [0.085, 0.1828] | [0.1683, 0.3634] | [0.0851, 0.1835] | [0.166, 0.36] | [0.0878, 0.1905] |

| Squared length (km) | 1.3434 | 1.5524 | 1.3563 | 1.5688 | 1.327 | 1.5194 |
|---|---|---|---|---|---|---|
| | [1.2008, 1.5031] | [1.3918, 1.7315] | [1.208, 1.5228] | [1.4005, 1.7572] | [1.1864, 1.4842] | [1.3618, 1.6954] |

Table 2 - continued

| Variable of interest | Time cost of public transports (z-score) | | Maximum distance to station (z-score) | | Mean areal education (%) | |
|---|---|---|---|---|---|---|
| | ST/SB | ST/PA | ST/SB | ST/SB | ST/SB | ST/PA |
| Intercept | 0.0068 | 0.007 | 0.0169 | 0.0169 | 0.0169 | 0.003 |
| | [0.0002, 0.1991] | [0, 0.02] | [0.0006, 0.5095] | [0.0006, 0.5095] | [0.0006, 0.5095] | [0.0004, 0.0889] |
| Variable of interest | 0.8034 | 0.7407 | 1.1775 | 1.1775 | 1.1775 | 1.2561 |
| | [0.6423, 1.0048] | [0.5952, 0.9218] | [1.0127, 1.3692] | [1.0127, 1.3692] | [1.0127, 1.3692] | [1.0816, 1.4588] |
| Squared variable of interest | | | 0.9985 | 0.9985 | 0.9985 | 0.9978 |
| | | | [0.9968, 1.0001] | [0.9968, 1.0001] | [0.9968, 1.0001] | [0.9961, 0.9994] |
| Reported importance of var. of interest | 1.2792 | 1.4618 | 0.7417 | 0.7417 | 0.7417 | 0.6586 |
| | [0.9146, 1.7891] | [1.0308, 2.073] | [0.5201, 1.0576] | [0.5201, 1.0576] | [0.5201, 1.0576] | [0.4545, 0.9543] |
| Age (years) | 0.9609 | 1.0341 | 1.0331 | 1.0331 | 1.0331 | 1.1448 |
| | [0.6685, 1.3812] | [0.7086, 1.509] | [0.7132, 1.4964] | [0.7132, 1.4964] | [0.7132, 1.4964] | [0.7771, 1.6866] |
| Sex (1 = male) | 0.7805 | 0.9885 | 0.6408 | 0.6408 | 0.6408 | 0.7545 |
| | [0.3729, 1.6335] | [0.4585, 2.1312] | [0.3039, 1.3512] | [0.3039, 1.3512] | [0.3039, 1.3512] | [0.3464, 1.6431] |
| Income category = 2 | 0.6832 | 0.66 | 0.5801 | 0.5801 | 0.5801 | 0.5305 |
| | [0.2769, 1.6858] | [0.2581, 1.6878] | [0.2354, 1.4296] | [0.2354, 1.4296] | [0.2354, 1.4296] | [0.207, 1.3595] |
| Income category = 3 | 0.7828 | 0.7781 | 0.6569 | 0.6569 | 0.6569 | 0.574 |
| | [0.3006, 2.0385] | [0.2876, 2.105] | [0.2507, 1.7217] | [0.2507, 1.7217] | [0.2507, 1.7217] | [0.2099, 1.57] |
| Education category = 2 | 2.0265 | 2.609 | 2.4498 | 2.4498 | 2.4498 | 3.4081 |
| | [0.6176, 6.6494] | [0.7578, 8.9824] | [0.7408, 8.101] | [0.7408, 8.101] | [0.7408, 8.101] | [0.9775, 11.8816] |
| Education category = 3 | 1.1646 | 1.5734 | 1.3773 | 1.3773 | 1.3773 | 1.9984 |
| | [0.5008, 2.7081] | [0.6537, 3.7875] | [0.5873, 3.2303] | [0.5873, 3.2303] | [0.5873, 3.2303] | [0.8204, 4.8677] |
| In couple (1 = yes) | 0.8384 | 0.6097 | 0.8862 | 0.8862 | 0.8862 | 0.6897 |
| | [0.3576, 1.9652] | [0.2513, 1.4791] | [0.3751, 2.0935] | [0.3751, 2.0935] | [0.3751, 2.0935] | [0.281, 1.693] |
| Children <14 at home | 1.2709 | 1.3977 | 1.5052 | 1.5052 | 1.5052 | 1.7936 |
| | [0.5765, 2.8016] | [0.6133, 3.1855] | [0.6922, 3.2732] | [0.6922, 3.2732] | [0.6922, 3.2732] | [0.7965, 4.0392] |
| Mean area education (%) | 1.1818 | 1.2751 | | | | |
| | [1.0181, 1.3718] | [1.0989, 1.4795] | | | | |
| Squared mean area education (%) | 0.9983 | 0.9975 | | | | |
| | [0.9967, 1] | [0.9959, 0.9991] | | | | |
| Length (km) | 0.2401 | 0.1227 | 0.2517 | 0.2517 | 0.2517 | 0.1255 |

|  | [0.1626, 0.3546] | [0.0831, 0.1811] | [0.1724, 0.3676] | [0.1724, 0.3676] | [0.1724, 0.3676] | [0.0862, 0.1829] |
| --- | --- | --- | --- | --- | --- | --- |
| Squared length (km) | 1.399 | 1.5636 | 1.3212 | 1.3212 | 1.3212 | 1.5313 |
|  | [1.2023, 1.6279] | [1.3478, 1.814] | [1.1827, 1.4759] | [1.1827, 1.4759] | [1.1827, 1.4759] | [1.3747, 1.7056] |

Table 3 : Predicted compositions (%) of physical activity behaviors within the ranges of observed values

|  | Green ratio | | | Destination density | | | Public transports time cost | | | Mean area education | | |
| --- | --- | --- | --- | --- | --- | --- | --- | --- | --- | --- | --- | --- |
| Percentile of variable of interest | Sedentary | ST | PA | Sedentary | ST | PA | Sedentary | ST | PA | Sedentary | ST | PA |
| Min. | 53.03 | 25.86 | 21.11 | 42.49 | 33.98 | 23.53 | 33.98 | 32.25 | 33.77 | 86.6 | 9.14 | 4.26 |
| 10 | 49.09 | 27.77 | 23.14 | 41.73 | 32.5 | 25.77 | 39.04 | 30.82 | 30.14 | 78.15 | 13.95 | 7.89 |
| 20 | 45.17 | 29.63 | 25.2 | 40.86 | 30.99 | 28.15 | 44.32 | 29.1 | 26.58 | 68.45 | 18.99 | 12.56 |
| 30 | 41.3 | 31.42 | 27.28 | 39.88 | 29.46 | 30.66 | 49.7 | 27.14 | 23.16 | 59.12 | 23.42 | 17.46 |
| 40 | 37.54 | 33.11 | 29.35 | 38.81 | 27.91 | 33.28 | 55.06 | 25.01 | 19.93 | 51.53 | 26.78 | 21.69 |
| 50 | 33.92 | 34.69 | 31.39 | 37.63 | 26.36 | 36.01 | 60.28 | 22.77 | 16.95 | 46.36 | 29.04 | 24.61 |
| 60 | 30.47 | 36.14 | 33.39 | 36.37 | 24.81 | 38.82 | 65.25 | 20.5 | 14.25 | 43.73 | 30.34 | 25.93 |
| 70 | 27.23 | 37.45 | 35.32 | 35.03 | 23.27 | 41.71 | 69.88 | 18.26 | 11.86 | 43.63 | 30.8 | 25.57 |
| 80 | 24.21 | 38.61 | 37.18 | 33.61 | 21.74 | 44.65 | 74.12 | 16.11 | 9.77 | 46.02 | 30.37 | 23.61 |
| 90 | 21.42 | 39.62 | 38.95 | 32.14 | 20.24 | 47.62 | 77.93 | 14.09 | 7.98 | 50.9 | 28.86 | 20.25 |
| Max. | 18.88 | 40.49 | 40.63 | 30.61 | 18.78 | 50.61 | 81.31 | 12.23 | 6.47 | 58.13 | 26.01 | 15.87 |

Table 4 : **Exponentiated coefficients (β) [95% confidence interval] of mixed linear regressions** modelling the relationship between personal and environmental variables and the log ratios ST/SB and PA/SB. All variables of interest with p-value<0.05 were included together in the same model.

|  |  | ST/SB | PA/SB |
|---|---|---|---|
|  | intercept | 0.0084 [0.0003, 0.2497] | 0.0011 [0, 0.0332] |
| Variables of interest | green ratio (%) | 1.1059 [1.0158, 1.2041] | 1.1209 [1.0286, 1.2215] |
|  | time cost of transports (z-score) | 0.8496 [0.6887, 1.0481] | 0.7954 [0.6449, 0.9811] |
|  | mean area education (%) | 1.1851 [1.0291, 1.3648] | 1.2785 [1.1084, 1.4747] |
|  | mean area education$^2$ (%) | 0.9983 [0.9967, 0.9998] | 0.9975 [0.9959, 0.9991] |
| Reported importance of variable in choice of neighborhood of residence | greenery | 0.9583 [0.6752, 1.3601] | 0.8598 [0.597, 1.2383] |
|  | SES | 0.7733 [0.5528, 1.0817] | 0.7134 [0.5028, 1.0121] |
|  | public transports | 1.2657 [0.914, 1.7528] | 1.4787 [1.0533, 2.076] |
| Personal variables | age (years) | 0.9839 [0.6966, 1.3895] | 1.0703 [0.7467, 1.5341] |
|  | sex (1 = male) | 0.7233 [0.3538, 1.4786] | 0.8139 [0.3864, 1.7144] |
|  | income category = 2 | 0.6258 [0.2688, 1.4571] | 0.5748 [0.2383, 1.3864] |
|  | income category = 3 | 0.6597 [0.2674, 1.6277] | 0.5956 [0.2325, 1.5261] |
|  | education category = 2 | 1.9415 [0.6382, 5.9067] | 2.4726 [0.7755, 7.8836] |
|  | education category = 3 | 1.2032 [0.5434, 2.6641] | 1.574 [0.6873, 3.6047] |
|  | in couple (1 = yes) | 0.8899 [0.3975, 1.9922] | 0.6727 [0.2905, 1.558] |
|  | children < 14 at home | 1.3388 [0.6324, 2.8343] | 1.5356 [0.7023, 3.358] |
| Trip caracteristics | length (km) | 0.2538 [0.1752, 0.3678] | 0.1363 [0.0936, 0.1985] |
|  | length$^2$ (km) | 1.3929 [1.2083, 1.6057] | 1.5578 [1.3512, 1.796] |

# Abbreviations

GPS: Global Positioning System; PA: Physical Activity; SB: sedentary behavior; ST: Standing or light activity; SES: Socio-economic Status; SD: standard deviation.

# Declarations

### Ethics approval and consent to participate

All participants signed an informed consent form. The study protocol was approved by the French Data Protection Authority (Decision No. DR-2013-568 on 2/12/2013).

### Consent for publication

All authors have read and approved the final manuscript and are in agreement with this submission.

### Availability of data and material

Data are available upon request, in the context of scientific collaboration.

### Competing interests

The authors declare no conflicts of interest.

### Funding

The RECORD MultiSensor Study was supported by the Ministry of Ecology (DGITM); Cerema (Centre for the Study of and Expertise on Risks, the Environment, Mobility, and Planning); INPES (National Institute for Prevention and Health Education); STIF (Ile-de-France Transportation Authority); and DRIEA (Regional and Interdepartmental Direction of Equipment and Planning) of Ile-de-France.

**Authors' contributions:** ID participated in the design of the study, analyzed and interpreted the results, and wrote the manuscript. BC participated in the design of the present study, and is the principal investigator of the RECORD MultiSensor Study. CS, AB, and JV provided critical revisions for intellectual content.

### Acknowledgements

NA

# Figures

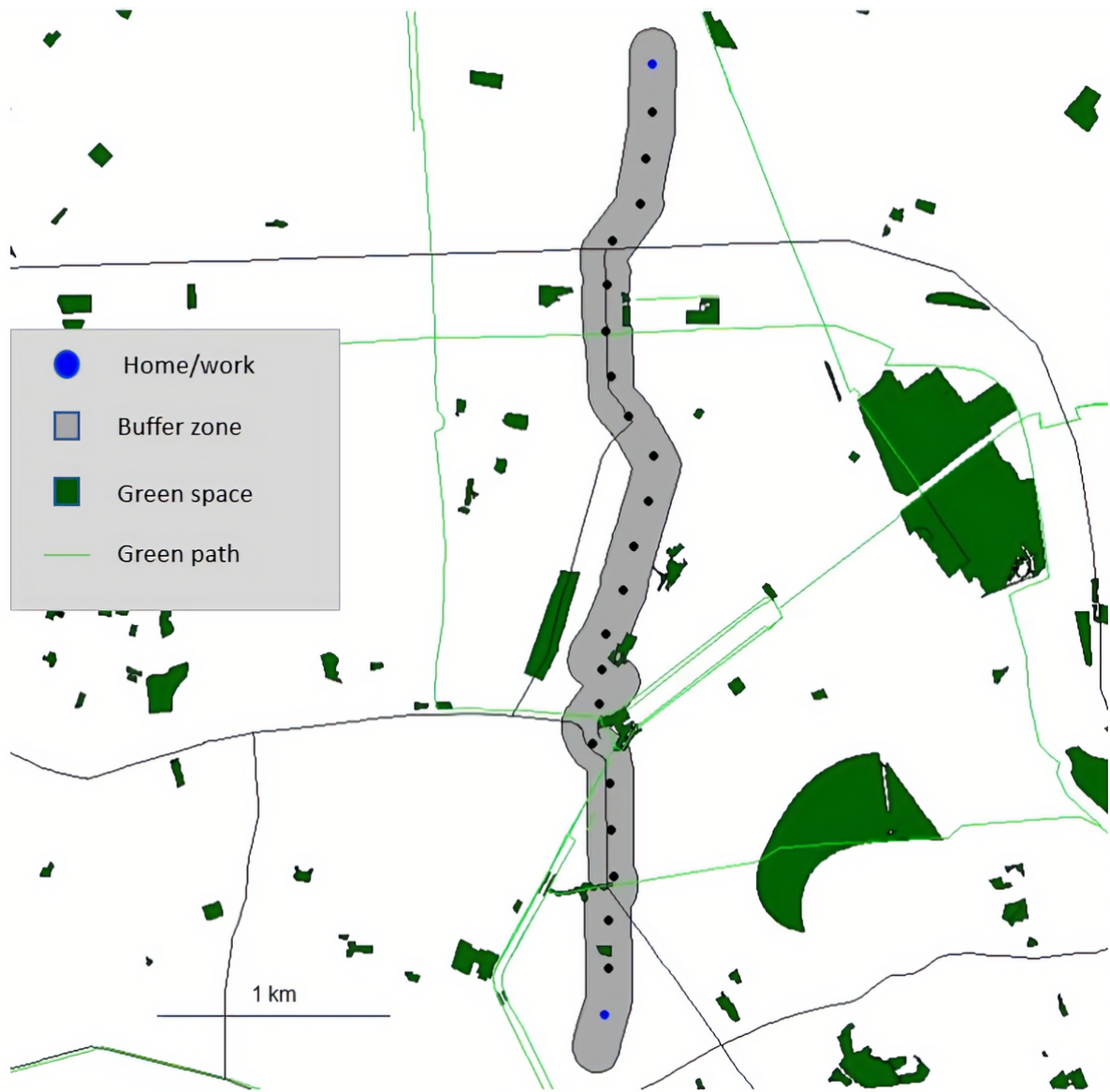

Figure 1

Examples of the shortest (walking) route of a work-home journey in the North of Paris and the neighboring Aubervilliers. The green ratio is the ratio of the green areas intersecting with the buffer along the route to the total buffer area. The distance to the network of green roads is the average of the distances between the points sample along the route.

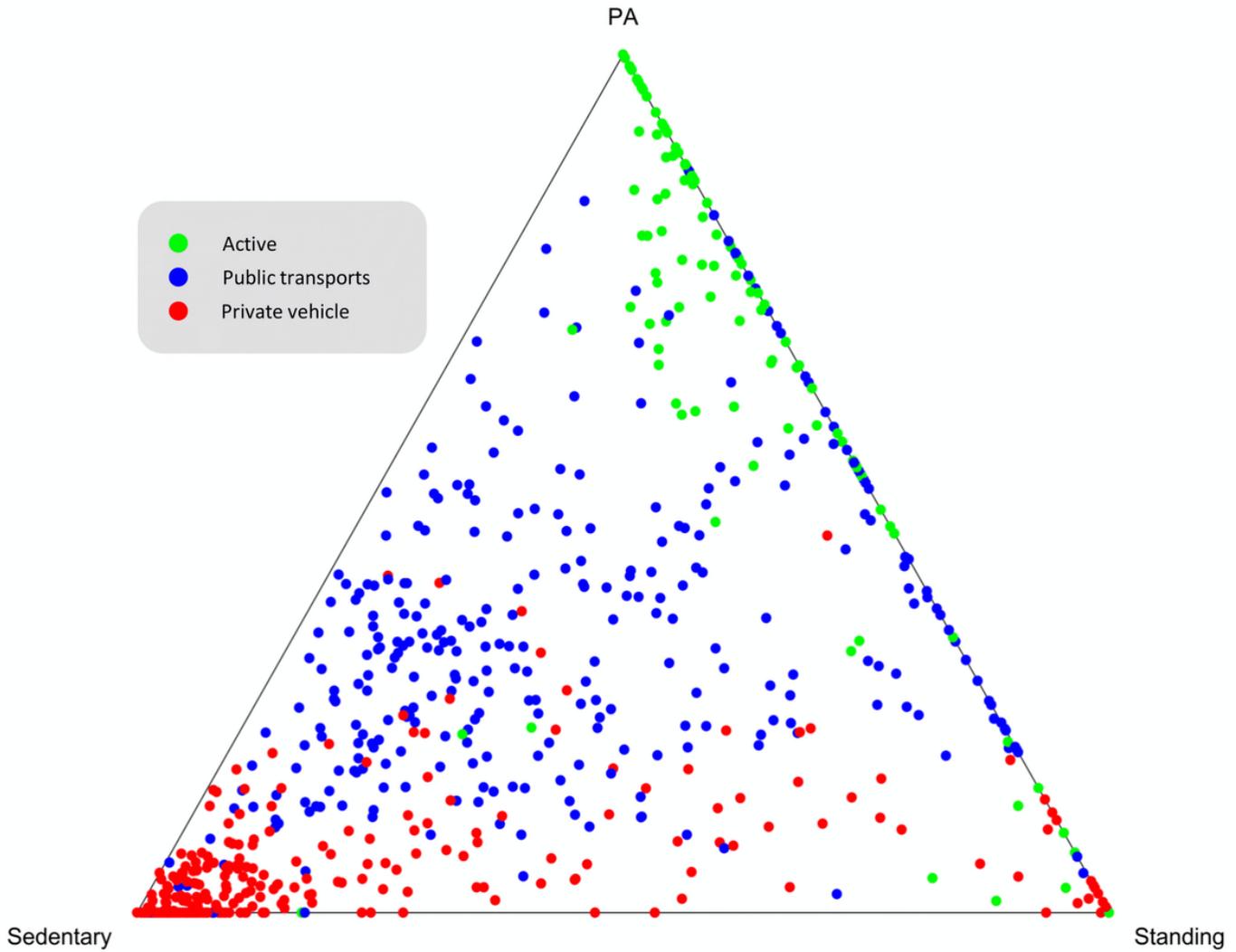

### Figure 2

Time-budget of physical activity behaviors over the journey by transportation mode. Colors indicate the transportation mode: active - all the journey was made using physically active modes (typically walking or bicycling); public transports - public transports were used at least partly during the journey, and private motorized vehicle.

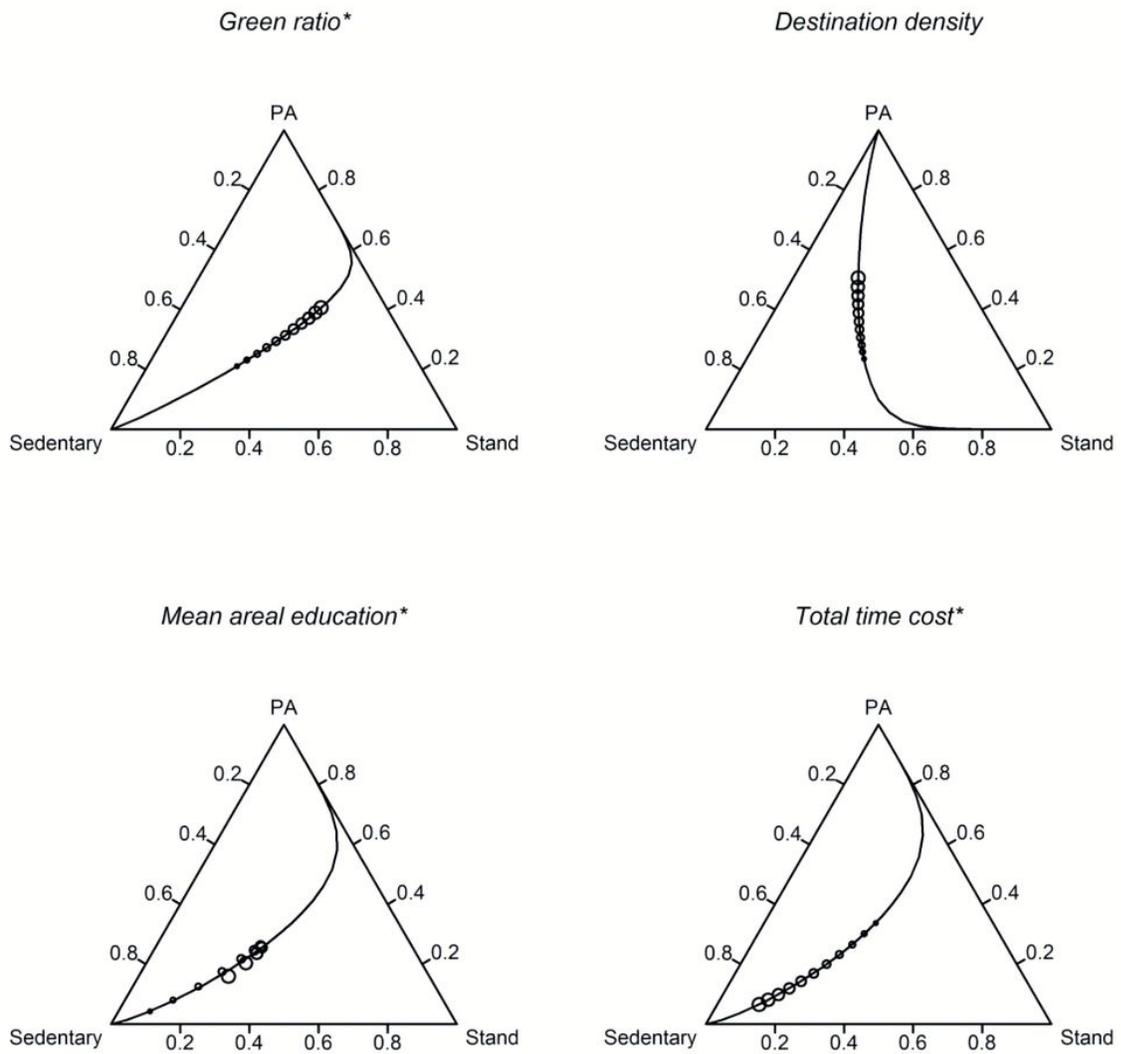

### Figure 3

Predicted time-budgets of physical activity behaviors for various values of the environmental variable of interest while all covariables take their mean values. The circles, from the smallest to the biggest, represent the predicted compositions for the lowest to the highest value of the explanatory variable in the population (by an increase of one decile).